\documentclass[10pt,twocolumn]{article}
\usepackage{ol2}
\usepackage{graphicx}
\usepackage{amsmath}

\begin{document}

\twocolumn[ 

\title{Optimal Alignment Sensing of a Readout Mode Cleaner Cavity}

\author{N. Smith-Lefebvre,$^{1}$  S. Ballmer,$^{2}$ M. Evans,$^{1}$ S. Waldman,$^{1}$ K. Kawabe,$^{3}$ V. Frolov,$^{4}$ N. Mavalvala$^{1}$}

\address{
$^1$LIGO Laboratory, Massachusetts Institute of Technology, Cambridge, MA 02139.\\
$^2$Syracuse University, Syracuse, New York 12344.\\
$^3$LIGO Hanford Observatory, PO Box 159, Richland, WA 99352-0159.\\
$^4$LIGO Livingston Observatory, PO Box 940, Livingston, LA 70754-0940.
}
\begin{abstract}
Critically coupled resonant optical cavities are often used as mode cleaners in optical systems to improve the signal to noise ratio (SNR) of a signal that is encoded as an amplitude modulation of a laser beam. Achieving the best SNR requires maintaining the alignment of the mode cleaner relative to the laser beam on which the signal is encoded.  An automatic alignment system which is primarily sensitive to the carrier field component of the beam will not, in general, provide optimal SNR. We present an approach that modifies traditional dither alignment sensing by applying a large amplitude modulation on the signal field, thereby producing error signals that are sensitive to the signal sideband field alignment. When used in conjunction with alignment actuators, this approach can improve the detected SNR; we demonstrate a factor of 3 improvement in the SNR of a kilometer-scale detector of the Laser Interferometer Gravitational-wave Observatory. This approach can be generalized to other types of alignment sensors. 
\end{abstract}
 ] 


In any quantum noise limited photodetection process where some physical quantity is encoded as an amplitude modulation on a laser beam, higher-order modes (HOM) of the transverse electromagnetic (TEM) optical field present a problem. Often the HOM carry no signal information, but nevertheless contribute to the noise. Optical sensing systems where HOMs can compromise the signal to noise ratio (SNR) include laser interferometer gravitational wave detectors \cite{LIGO, Virgo, GEO600}, cavity optomechanics \cite{kippenbergScience2008} experiments, and quantum optics experiments \cite{QO}.

A critically coupled optical cavity can attenuate the HOM content of a laser beam. Such cavities also act as temporal filters for reducing laser amplitude and phase noise for frequencies above the cavity linewidth. When such mode cleaner filter cavities are placed at the output (readout) port of an optical system, deriving error signals to control the length and alignment of the Output Mode Cleaner (OMC) cavity poses a particular challenge. The signal-rich optical field may be weak compared to the HOM components. 

In this article we describe a solution to the problem of aligning the OMC cavity used at the output port of the 4 kilometer long laser interferometers of the Laser Interferometer Gravitational-wave Observatory (LIGO)\cite{LIGO}. Though we consider the LIGO optical readout here, our scheme applies to any optical system where a signal is encoded as an amplitude modulation of a laser field, and the spatial mode of the signal-induced modulation sideband differs from that of the DC carrier field.  In the case of LIGO, the signal is a gravitational wave induced modulation of the optical field, typically at frequencies 10 Hz to 10 kHz shifted from the carrier. 

\begin{figure}
  \begin{center}
  \leavevmode
  \includegraphics[width=0.45\textwidth]{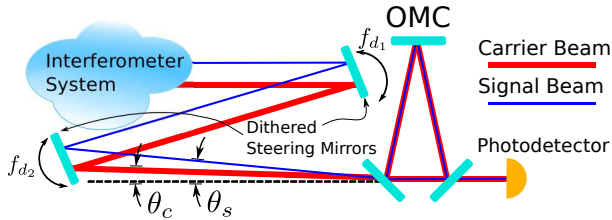}
  \end{center}
  \caption{A signal is encoded by interferometry (or any other means) as an amplitude modulation of a laser beam. The beam is then steered by two mirrors which are dithered in angle before passing though a mode cleaner cavity. Alignment signals can be derived by demodulation of the transmitted photocurrent. The misalignment has been grossly exaggerated in the figure.}
  \label{fig:blockdiag}
\end{figure}

%
Fig. \ref{fig:blockdiag} shows a schematic representation of the readout system we consider. The laser beam incident on the OMC comprises a carrier field and signal sideband fields which must be aligned to the OMC cavity. In the absence of technical noise sources, an automatic alignment system should maintain the cavity alignment that maximizes the SNR of the detected signal with respect to the photon shot noise. In general the carrier and sideband fields do not occupy the same spatial modes and their transmission through the OMC varies differently as a function of the input alignment. 

As a simple example, consider a beam reflected from an over-coupled Fabry-Perot cavity (OC). The OC is held slightly offset from resonance such that periodic length excitations cause amplitude modulation that can be detected on a photodetector. Consider the case of a static misalignment of the input carrier field. The reflected carrier field will also be misaligned, while the signal field, being generated inside the OC, will be in the same spatial mode as the OC. Thus the carrier and signal fields will have a relative misalignment. In this example, the cause of the alignment mismatch is a misalignment of the input field to the OC. However, in more complicated cases, such as in the case of a LIGO detector composed of multiple, coupled cavities, the source of extra modal content may not be so easily removed.

Consider a carrier field with transmitted amplitude $c$ and amplitude modulated upper and lower signal sidebands fields with transmitted amplitudes $s$. The shot noise limited SNR of the signal when measured on a photodetector is proportional to
\begin{equation}
SNR \propto \frac{cs}{\sqrt{c^2 + 2s^2}} \approx s.
\end{equation}
Here we have made the following simplifying assumptions: the carrier field amplitude is much greater than that of the signal field, the signal fields are pure amplitude modulation, and that there is perfect spatial overlap of the carrier and the sidebands after transmission through the OMC. We see that the alignment system that maximizes optical SNR of fields detected after transmission through an OMC simply maximizes the amplitude transmission of the {\it signal field} through the OMC. We do not address contributions to readout noise due to alignment jitter of the beam incident on the OMC in this article. 

To leading order, typical alignment schemes sense the ${\rm TEM}_{01}$ and ${\rm TEM}_{10}$  modes of the carrier field measured in the the OMC eigenmode basis\cite{Anderson1984}. The presence of these modes is interpreted as a misalignment. In this article we will show how to modify a common alignment technique to be sensitive to signal field misalignments, and thus allow one to maximize the SNR of the signal.

Consider a single alignment degree of freedom where misalignments of the carrier and the signal fields are represented by one set of misalignment angles $\theta_c$ and $\theta_s$. Dither alignment sensing is commonly achieved by mechano-optical modulation of an angular degree of freedom, e.g by driving steering mirrors that direct the beam into the OMC, as shown in Fig. \ref{fig:blockdiag}. Each mirror is dithered in angle with constant amplitude at a given frequency to modulate the input pointing into the cavity. 

\begin{figure}
  \begin{center}
  \leavevmode
  \includegraphics[width=0.45\textwidth]{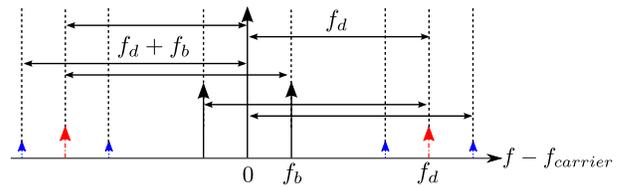}
  \end{center}
  \caption{Arrow diagram showing electric fields after transmission through the OMC. In the figure, $f_d$ is the angular dither frequency, $f_b$ is the beacon modulation frequency.}
  \label{fig:ditherarrows}
\end{figure}

Figure \ref{fig:ditherarrows} shows the frequency content of the field transmitted through the OMC. The carrier field is separated in frequency from two amplitude modulated signal sideband fields at $\pm f_b$. The dithering steering mirror produce two sidebands at $\pm f_d$, with a field amplitude proportional to $\theta_c$ (alternatively, the ${\rm TEM}_{01}$ mode amplitude\cite{siggJOSA2000}). The signal field also has dither sidebands proportional to $\theta_s$. The demodulated signal is made of products of field amplitudes separated by the demodulation frequency.

We will use the notation $P(f)$ to represent the OMC transmitted photocurrent demodulated at the frequency $f$. We define the standard dither alignment signal as
\begin{equation}
\label{eq:sstandard}
S_{standard} = P(f_d) \propto 2c^2 \theta_c + 4 s^2 \theta_s \approx 2c^2 \theta_c,
\end{equation}
where $c$ is the carrier field amplitude, $s$ is the signal field amplitude. For the case where the carrier power is much greater than the signal field power ($c^2 \gg s^2$), the demodulated alignment signal is sensitive only to misalignments of the carrier field. {\it In this scheme the error signal is nulled at an alignment where the total transmitted cavity power is maximized.} We will refer to this as the {\it standard dither scheme}.

%


To better sense the alignment of the signal field one may create a large amplitude modulation of the signal, say by modulating the length of the signal cavity. This modulation creates a frequency tag of the spatial mode of the signal field. We refer to this as a ``beacon'' modulation, and $f_b$ is the modulation frequency.

As in the standard dither scheme, steering mirrors are used to dither the angle of the beam. The desired error signal in this scheme is produced by demodulating the transmitted power at $f_b+f_d$ or $f_d-f_b$, or the sum of these signals. We define the beacon alignment signal as
\begin{equation}
\label{eq:sbeacon}
S_{beacon}=P(f_d+f_b) \propto 2sc\theta_c+2cs\theta_s.
\end{equation}

When this error signal is nulled, the SNR is improved in comparison to standard dither, but sensitivity to the carrier field remains, so the SNR will not be maximized. Because the beacon modulation is small, $S_{beacon}$ will have a higher relative noise level than the standard dither approach.

%

The beacon alignment scheme was compared directly to a standard dither scheme using the OMC cavity at the output of the 4 km LIGO Interferometer at Hanford (H1). The readout was arranged as in Fig. \ref{fig:blockdiag}. The angular dither frequencies were between 1.5 and 2.5 kHz, while the beacon modulation was a 10 Hz excitation of the differential arm length. 

The differential arm length of the interferometer is sensed as an amplitude modulation at the output port. The static carrier at the antisymmetric port is generated by an offset of the differential arm length. The OMC filters the spatial and frequency content of the beam before the beam is split equally and detected on two photodetectors. The LIGO interferometers are examples of optical systems where a beacon based alignment system performs significantly better than a standard dither scheme. 

\begin{figure}
  \begin{center}
  \leavevmode
  \includegraphics[width=0.45\textwidth]{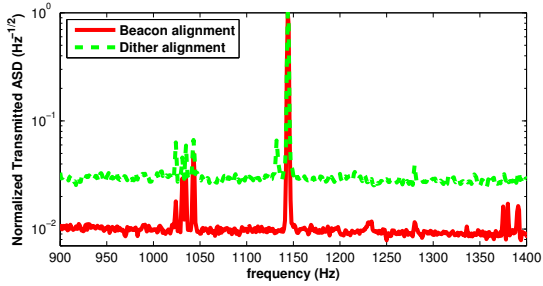}
  \end{center}
  \caption{The noise amplitude spectral density of the LIGO H1 detector using two types of alignment schemes. The curves are normalized to a calibration line at 1144 Hz.  The small line structures are resonances of the suspension wires supporting the mirrors. The beacon scheme shows an SNR improvement of about a factor of 3.}
  \label{fig:shotnoise}
\end{figure}

Fig. \ref{fig:shotnoise} shows the amplitude spectral density of the transmission of the OMC of the H1 interferometer. The plot is centered on an injected calibration signal at 1144 Hz and the curves are normalized to the peak height of this line. The calibration line is surrounded by primarily shot noise. The dashed green curve (color online) shows the performance using dither sensing, while the solid red curve shows beacon sensing. The SNR of the calibration line is improved by about a factor of 3.1 by using a beacon scheme. A factor of 2.4 is due to an increase of signal strength and the remaining factor of 1.3 is due to a reduction in the total transmitted power, and hence the shot noise. We propose that the poor performance of the standard technique is due to excess HOMs in addition to the ${\rm TEM}_{01}$ mode associated with misalignment.

It is possible to combine the standard dither signal with the beacon signal to produce a signal which is sensitive to signal field misalignments only. We also make use of the DC transmitted power, $P_{DC} = P(0) \approx c^2$, and the beacon modulation amplitude, $P(f_b) \approx 2cs$. An optimal alignment signal can be constructed as follows:
\begin{align}
\label{eq:optsig}
S_{optimal} &= S_{beacon} - \tfrac{P(f_b)}{2P_{DC}} S_{standard}\\
\nonumber &\propto 2cs\theta_s+2sc\theta_c-\tfrac{2cs}{2c^2}(2c^2\theta_c) \\
\nonumber &\propto 2cs\theta_s.
\end{align}

An alternative (though mathematically equivalent) technique would be to just make a dither servo which maximizes the optical SNR directly. This may be a more desirable approach depending on how signals are read out and if digital processing is possible. In this approach, the SNR should be calculated in real time and demodulated at the dither frequency to provide the error signal.

As above, if the beacon amplitude on the photodetector is $P(f_b)$ and the DC power is $P_{DC}$, the optical SNR is
\begin{equation}
SNR = \tfrac{P(f_b)}{\sqrt{P_{DC}}}.
\end{equation}
Any dither sensing scheme is essentially measuring the partial derivative of a signal with respect to the dithered degree of freedom \cite{Kawabe:94}. Thus, a dither servo which maximizes the SNR will measure a signal proportional to
\begin{equation}
\tfrac{\partial}{\partial \theta} \left( \tfrac{P(f_b)}{\sqrt {P_{DC}} } \right) = \tfrac{1}{\sqrt {P_{DC}}} \left( \tfrac{\partial P(f_b)}{\partial \theta} - \tfrac{P(f_b)}{2P_{DC}} \tfrac{\partial P_{DC}}{\partial \theta} \right),
\end{equation}
which is equivalent to (\ref{eq:optsig}) up to constant factors. This also shows that this signal is nulled at maximum SNR.

In gravitational-wave detectors, the OMC transmission is often held constant by a control system. If $f_d$ is within the control bandwith, but $f_b$ is not, then the carrier alignment sidebands will be suppressed. Sufficient suppression makes the beacon scheme approximately equivalent to the optimal scheme. This was the configuration used by both the L1 and H1 interferometers in the sixth LIGO science run \cite{TobinDC}.


The technique of using a beacon modulation to sense misalignment of the signal field can be generalized to other types of alignment sensors, for example: split quadrant photodetectors detecting light picked off the beam entering the OMC or split quadrant detectors in reflection of the OMC coupled with frequency or length dithered sidebands, also known as wavefront sensors.

We define the standard carrier alignment signal as $G_{standard}$. We also define another signal, demodulated $f_b$ away from the standard signal, as $G_{beacon}$. $P_{DC}$ and $P(f_b)$ are defined as in (\ref{eq:optsig}). The generalized signal is:
\begin{equation}
G_{optimal} = G_{beacon} - \tfrac{P(f_b)}{2P_{DC}} G_{standard}.
\end{equation}
We emphasize that the $G$ signals are derived from the alignment sensor, while the $P$ signals are derived from the OMC transmission.

To summarize, we have introduced the concept of using a beacon as a strong modulation close to signal frequencies to generate alignment signals that lead to increased SNR compared to the standard dither scheme. The beacon scheme demonstrates a factor of 3 improvement in SNR for the LIGO H1 detector. Finally, we propose a detection scheme to give optimum SNR when used to align an OMC or filter cavity.

We gratefully acknowledge illuminating discussions with Tobin Fricke, Jeff Kissel and Daniel Sigg. LIGO was constructed by the California Institute of Technology and Massachusetts Institute of Technology with funding from the National Science Foundation and operates under cooperative agreement PHY-0757058. This paper has LIGO Document Number LIGO-P1100025.



\end{document}